\documentclass{jetpl}

\twocolumn \lat \RequirePackage{cite}

\DeclareMathOperator{\Tc}{T_c}

\DeclareMathOperator{\sign}{sign}

\DeclareMathOperator{\sym}{Sym}

\title{Quantum and Classical Binomial Distributions for the
Charge Transmitted through Coherent Conductor.}

\rtitle{Quantum and Classical Binomial Distributions\ldots}

\sodtitle{Quantum and Classical Binomial Distributions for the
Charge Transmitted through Coherent Conductor.}

\author{G.\,B.\,Lesovik$^{\dagger}$\/\thanks{lesovik@landau.ac.ru},
N.\,M.\,Chtchelkatchev$^{\dagger,\ddagger}$}

\rauthor{G.\,B.\,Lesovik, N.\,M.\,Chtchelkatchev}

\sodauthor{Lesovik, Chtchelkatchev}

\address{$^\dagger$L.D.\ Landau Institute for Theoretical Physics,
Russian Academy of Sciences, 117940 Moscow, Russia.}
\address{$^\ddagger$Institute for High Pressure Physics, Russian Academy
of Sciences, Troitsk 142092, Moscow Region, Russia}
\dates{}{*}

\abstract{We discuss controversial results for the statistics of
charge transport through coherent conductors. Two distribution
functions for the charge transmitted was obtained previously,
one actually coincides with
classical binomial distribution the other  is different and we call it
here quantum binomial distribution.  We show, that high order charge
correlators, determined by the either distribution functions, can
all be measured in different setups. The high order current
correlators, starting the third order, reveal (missed in previous
studies) special oscillating frequency dependence on the scale of
the inverted time flight from the obstacle to the measuring point.
Depending on setup, the oscillating terms give substantially
different contributions.}


\begin{document}

\maketitle

Last years has appeared new direction in quantum transport
investigations ---  description of the statistics of a charge
transmitted through a quantum conductor. Usually the distribution
functions are investigated for a charge $Q _ {t_0}$, transmitted
during a large interval of time $t_0$ through a certain
cross-section of the conductor, and, that is essential, all
observables are calculated from the first principles. Despite of
appreciable amount of articles (see the review \cite {levitov})
and received results, some questions, in particular concerning the
measurement theory, still remain unclear. One of the problems is
that in a quantum case (in contrast to classical) arises the
question how to define the observable that should be calculated.
Technically this uncertainty is connected with noncommutativity of
the current operators at various times. As it appears, it is
possible to present several definitions for the distribution
function (DF) and characteristic function (CF) which a) in the
classical case coincide, b) satisfy some general principles (in
particular, correlators $ \langle Q _ {t_0} ^n \rangle$ prove to
be real), however lead to different answers in quantum case. It is
possible to understand what definition is ``correct''  only having
analyzed definite set-up for (at least gedanken) measurement.

In this paper we shall consider few variants of measurements and
corresponding definitions of  CF.  We shall also comment the
results received earlier. In the first paper devoted to
microscopic description of the distribution function \cite {LL1}
the following definition was accepted:
\begin{gather}
\chi(\lambda) = \left\langle \exp\{i\lambda \hat
Q(t_0)\}\right\rangle=\left\langle \exp\left\{i\lambda
\int_0^{t_0} dt\hat I(t)\right\}\right\rangle \label{chiLL1}
\end{gather}
(here $ \langle\ldots \rangle $ denotes ensemble averaging). This
definition is the most direct generalization of the classical
definition; it differs only in the replacement of the charge and
current observables by the appropriate operators. Performing
calculations we considered  large intervals of time at which the
correlators $ \langle \langle \hat Q(t_0)^n\rangle \rangle =
\langle \langle \int_0^{t_0}dt_1...\int_0^{t_0} dt_n \hat
I(t_1)\ldots\hat I(t_n) \rangle \rangle $ approximately equal to
\begin{gather}
\langle \langle \hat Q(t_0)^n\rangle \rangle = t_0\langle \langle
\hat I_0^n \rangle \rangle, \label{QnInt}
\end{gather}
where $\langle \langle\hat I^n \rangle \rangle_{0}$ is the
irreducible current correlator of $n$-th order at zero frequency
limit. (The irreducible correlators satisfy the equation, $
\langle \exp\{i\lambda \hat Q(t)\}\rangle=
\exp\{\sum_{n=1}^{\infty}(i\lambda)^n \langle \langle\hat
Q(t)^n\rangle \rangle/n!\}$.)

The method of calculation used in \cite {LL1} can be generalized
for the case of finite temperatures and (for normal single-channel
conductors) we find
\begin{multline}
 \chi (\lambda)=
\exp\biggl\{ t_0 g\int \frac{d\epsilon}{2\pi
 \hbar} \ln\{(1-n_L)(1-n_R) +n_Ln_R +
 \\
n_L(1-n_R)\chi_{\epsilon} (\lambda) +
  n_R(1-n_L)\chi_{\epsilon} (-\lambda)\}\},
\label{chigeneral}
\end{multline}
where $\chi_{\epsilon}(\lambda)= \cos\{\lambda
e\sqrt{T(\epsilon)}\} +i \sqrt{T(\epsilon)}\sin\{\lambda
e\sqrt{T(\epsilon)}\}$, $g=2$ is the factor taking into account
spin degeneracy, $T$  is the transparency, and $n_{L, R}$ are  the
filling factors in in the left and right reservoirs
correspondingly. This distribution (at $k_BT=0, eV\neq0 $) we
shall name {\it quantum} binomial distribution. In multichannel
case $ \chi (\lambda) $ is a product $ \Pi_n \chi_n (\lambda) $ of
the characteristic functions $ \chi_n (\lambda) $ corresponding to
transmission eigenvalues $T_n$. The distribution function  (\ref
{chigeneral}) formally  describes fractional charge transport. At
indirect measurements of a charge in solid-state systems the
fractional charge may in principle appear, for example, in the
shot noise in conditions of fractional quantum Hall effect \cite
{frac}. In our case the size of a ``of a charge quantum'' $2e\sqrt
{T} $ is determined by the eigen-value of the current operator
provided that its action is considered on the subspace of
one-particle excitations at the given energy $e\sqrt {T} = \pm
\langle \pm |\hat I (t_0) | \pm \rangle$,  where the normalized
one-particle excitations $ | \pm \rangle$  satisfy the condition $
\langle + | \hat I (t_0) | - \rangle =0$,  see also \cite {LL1}.
If we take into account the logarithmic on time $t_0$ corrections
to the irreducible correlators then the exact charge quantization
which follows from  the discrete distribution function  is
replaced apparently by small modulations in the continuous
distribution function. If we limit ourself to the corrections
(occurring from the vacuum fluctuations) to the pair correlator
then the distribution function becomes
\begin{multline*}
P(Q)=\sum_n
P^{(0)}(ne\sqrt{T})\left(\frac{2\pi}{G\hbar\ln\{t_0\omega\}}\right)^{1/2}\times
\\
\exp\{-(Q-ne\sqrt{T})^2/ 2G \hbar\ln\{t_0\omega\} \},
\end{multline*}
where $P^{(0)}(ne\sqrt{T})$ is the discrete distribution function
with disregarded logarithmic corrections, $G$ is  the conductance,
$\omega$ is a characteristic frequency scale of the conductance
dispersion.

Using Eq.~(\ref{chigeneral}) we find for the third order
correlator:
\begin{multline*}
\langle \langle(\hat Q_{t_0}^3)\rangle \rangle= - t_0  g\int
\frac{d\epsilon}{2\pi \hbar}e^3 T^2(n_L- n_R)\times
\\
\left[\{3[n_L(1-n_R) + n_R(1-n_L)]-1\}- 2T(n_L- n_R)^2\right].
\end{multline*}

At small voltage (and when the transparency $T $ does not depend
on energy) the correlator is proportional to $V^3 $ and at large
$V $ we get \cite{LL1}:
\begin{equation}
\langle \langle Q_{t_0}^3\rangle \rangle =
-2e^3T^2(1-T)\frac{2eVt_0}{h}. \label{QQQ1}
\end{equation}
According to Eq.~(\ref {QnInt}) the third order current correlator
at zero frequencies  is
\begin{equation}
\langle \langle I_{0}^3\rangle \rangle
=-2e^3T^2(1-T)\frac{2eV}{h}. \label{III1}
\end{equation}
Since it was not quite clear how to measure \textit{quantum}
binomial DF and correlators as in Eq.~(\ref {QQQ1}), it was
suggested in Ref.~\cite {LLL} to use a spin located near to a wire
as the counter of electrons passed through it. Thus it appears
that the definition for  CF in this case differs from (\ref
{chiLL1}) by the presence of the time ordering:
\begin{equation}
\chi(\lambda) = \langle \tilde T  e^{i\lambda/2 \int_0^{t_0} \hat
I(t)dt } Te^{i\lambda/2 \int_0^{t_0} \hat I(t)dt } \rangle,
\label{chiLL2}
\end{equation}
in this (and only this) formula  the symbol $T$ means the usual
time ordering, and $ \tilde{ T} $ means the ordering in the
opposite direction.

It was found in \cite {LLL, LL2} (see also references in \cite
{levitov}) for the third order correlator of the transmitted
charge
\begin{equation}
\langle \langle Q_{t_0}^3\rangle \rangle =
e^3T(1-T)(1-2T)\frac{2eVt_0}{h}. \label{QQQ2}
\end{equation}
As we see, the correlators of the third order (\ref {QQQ1}) and
(\ref {QQQ2}) essentially differ. It would seem that there is
nothing unexpected in such distinction as the definitions (\ref
{chiLL1}) and (\ref {chiLL2}) differ. However,  for example, the
third order correlator of a charge according to both definitions
(\ref {chiLL1}) and (\ref {chiLL2}) actually contains the
correlator of currents at small frequencies; but such current
correlator can be calculated with the help of the first definition
(\ref {chiLL1}) correctly at  zero frequency limit (\ref {III1})
and it can be checked independently using the same machinery that
was used in \cite {lesovik} for the calculation of the pair
correlator \cite {generation}. It appears that the dispersion of
the third order current correlator at small frequencies (along
with the difference of the definitions) results in different
answeres for the third order correlators. Really, at frequencies
$\omega \ll eV/\hbar $  and $x_{1,2,3} > 0 $ we have
\begin{multline}\label{Ix1Ix2Ix3}
\langle \langle I_{\omega_1}(x_1)I_{\omega_2}(x_2)
I_{\omega_3}(x_3)\rangle \rangle
=2\pi\delta(\omega_1+\omega_2+\omega_3)\times
\\
T(1-T)e^{i\omega_1 x_1/v_F
+i\omega_2x_2/v_F+i\omega_3x_3/v_F}\times
\\
[1-2T-e^{-2i\omega_2x_2/v_F}] eV\frac{2e^3}{h}.
\end{multline}
When $x_1=x_2=x_3=x$
\begin{multline}\label{IxIxIx}
\langle \langle I_{\omega_1}(x)I_{\omega_2}(x)
I_{\omega_3}(x)\rangle \rangle
=2\pi\delta(\omega_1+\omega_2+\omega_3)\times
\\
 T(1-T)[1-2T-e^{-2i\omega_2x/v_F}]eV\frac{2e^3}{h}.
\end{multline}
Formally assuming  that such dependence on frequency is correct at
all frequencies we find for the correlators in time
representation:
\begin{gather}\label{IxtIxtIxt}
\langle \langle I (t_1,x)I(t_2,x) I(t_3,x) \rangle \rangle
=eV\frac{2e^3}{h} T(1-T)\times
\\\notag
\sym\left([1-2T]\delta (t_i-t_j)\delta(t_j-t_k) - \delta (\tilde
t_i-\tilde t_j)\delta(\tilde t_j-\tilde t_k)\right),
\end{gather}
where the symbol $\sym$ means symmetrization on the indices $i\neq
j\neq k $; $ \tilde t_2\equiv t_2+2x/v_F $, $ \tilde t_1=t_1 $, $
\tilde t_3=t_3 $. (At the account of real frequency dependence
instead of $\delta$-functions there should stand functions which
decay on characteristic times $t\sim \tau_0$, \footnote
{Characteristic time scale for the decay of the correlators can be
estimated as $ \tau_0 \sim\hbar/eV$.  We plan to consider time
dependence of the correlators in detail in a separate article.}
but for simplicity we shall describe the case with
$\delta$-functions). Substituting this expression into the
expression for the third order charge correlator which follows
from (\ref {chiLL2}) we get
\begin{multline}
\langle \langle Q_{T}^3\rangle \rangle = \frac 3 4
\left[\int_0^Tdt_1\int_0^T dt_2\int_0^{t_2}dt_3 +\right.
\\
\left.\int_0^T dt_2\int_0^{t_2}dt_1\int_0^Tdt_3
+\int_0^Tdt_1\int_0^{t_1} dt_2\int_0^{t_2}dt_3+\right.
\\
\left.+  \int_0^Tdt_3\int_0^{t_3}dt_2\int_0^{t_2}dt_1\right]
\langle \langle I(t_1)I(t_2)I(t_3) \rangle \rangle \label{QQQTT}
\end{multline}
and integrating over times we find at final $x$ the answer (\ref
{QQQ2}) proportional to $T (1-T) (1-2T)$ that is typical for the
\textit{classical} binomial distribution (the same phenomenon
takes place for other irreducible high-order correlators).

It occurs because terms in the Eq.~\eqref{IxtIxtIxt} containing
$\delta$-functions that depend from $\tilde t_i$ do not give the
contribution to the answer as they are not equal to zero only when
simultaneously  $t_3>t_2$ and $t_1>t_2$ and the integration volume
in Eq.~(\ref {QQQTT}) does not cover such sector. Consider, e.g.,
the contribution to the correlator $\langle \langle Q_{T}^3\rangle
\rangle$ from the term in Eq.~\eqref{IxtIxtIxt} proportional to
\begin{gather}
\label{delta} \delta(\tilde t_1-\tilde t_3)\delta(\tilde
t_3-\tilde t_2)=\delta(t_1-t_3)\delta(t_3- t_2-2x/v_F).
\end{gather}
From Eq.~\eqref{delta} follows that the region where $t_1\approx
t_3\approx t_2+2x/v_F $ in Eq.~\eqref {QQQTT}  should give the
leading contribution to the integrals. But from the requirement
$x>0$ follows that $t_3>t_2$ and $t_1>t_2$. The region defined by
these inequalities does not overlap with the volume of the
integration in Eq.~\eqref{QQQTT}; thus the contribution
\eqref{delta} to $\langle\langle Q_{T}^3\rangle\rangle$  is equal
to zero. It is similarly possible to show that generally all terms
in Eq. \eqref {IxtIxtIxt}  proportional to $\sym\delta(\tilde
t_i-\tilde t_j)\delta(\tilde t_j-\tilde t_k)$ at  $x>0$ do not
give the contribution to $\langle\langle Q_{T}^3\rangle\rangle$.

One could say that  when the incident wave packet first fully
passes  the detector and only then, with a time delay, through the
detector goes back the part of the wave packet reflected from the
barrier, the specific quantum interference disappears and the
answer (\ref {QQQ2}) is true. But if the distance to the detector
is small then the incident wave packet interferes with the
reflected one in the measurement region that leads eventually to
the answer (\ref {QQQ1}). So Eq.~(\ref {QQQ2}) is true when the
time of electron flight from the scatterer to the spin of the
detector that is situated at the distance $d $ from the wire and
$L$ from the scatterer is larger than the decay time $ \tau_0 $ of
the correlators. If these requirements are violated the answer
will be different. In the calculations described in Ref. \cite
{LLL} though it was formally assumed  that the distance $L $ is
equal to zero, in fact was considered the limit when $L $ actually
exceeded the wave packet size (which was also kept to be zero).

For the case when the spin-detector is located near to the
scatterer $x\ll v_F\tau_0 $ and it is close to the wire $d\ll
v_F\tau_0 $, the answer for $ \langle \langle Q_{t_0} ^3\rangle
\rangle $ is proportional to $-T^2 (1-T) $ and it coincides with
the answer  (\ref {QQQ1}) obtained from the quantum distribution.
Really, using the general expression for the correlator
(\ref{Ix1Ix2Ix3}) at $x_{1,2,3}\ll v_F \tau_0$ we get
\begin{multline}
\langle \langle I_{\omega_1}(x_1)I_{\omega_2}(x_2)
I_{\omega_3}(x_3)\rangle \rangle \simeq
\\
-2\pi\delta(\omega_1+\omega_2+\omega_3)2T^2(1-T) eV\frac{2e^3}{h}.
\end{multline}
Using this expression at $ \omega \ll eV/\hbar $ and the
definition (\ref {QQQTT}) we obtain the expression for $ \langle
\langle Q_{t_0}^3\rangle \rangle $ proportional to $-T^2 (1-T) $
as well as in the calculation with the use of  CF (\ref {chiLL1}).

The measurement with the spin basically may be implemented in
practice with the help of muons which can be trapped near to the
conductor and then the measurement of the direction of their decay
would give the angle of their spin rotation in a magnetic field.
As an additional example of the measuring procedure which
basically can be implemented practically, we have analyzed the
measurement of the irreducible charge correlators with the help of
a ammeter represented by a semiclassical system (for example, an
oscillatory circuit) weakly  interacting with the current in a
quantum conductor. The state of the ammeter is characterized by
the magnitude $\phi$. Interaction of the ammeter with a quantum
conductor is described by the interaction Hamiltonian  $H_i
=\lambda \phi \hat I (t) $, where $ \lambda $ is the interaction
constant, $ \hat I (t) $ is the current operator in the quantum
conductor, $ \hat I (t) = \int \hat I (t, x) f (x) dx $ (we do not
take into account effects of retardation), thus the area of the
integration is determined by some kernel $f(x)$. Correlators $
\phi $ are expressed through correlators of currents in a quantum
conductor as follows:
\begin{multline}
\langle\langle (\phi(t))^n\rangle\rangle = \left(-\frac\lambda
2\right)^n\int_c d\tau_1\ldots d\tau_n
\\
\kappa(|t-\tau_1|)\sign(t-\tau_1)\ldots\kappa(|t-\tau_n|)
\sign(t-\tau_n)\times
\\
\langle\langle \Tc I(\tau_1)\ldots I(\tau_n) \rangle\rangle,
\end{multline}
where the integration is performed along the usual Keldysh
contour; $ \kappa (\tau) $ is the susceptibility of the ammeter.
In that specific case when the ammeter represents an oscillator,
the equation of motion for $ \phi $ is: $ \ddot {\phi} + \gamma \dot
{\phi} + \Omega^2 \phi =\lambda I (t)/M $; the susceptibility $
\kappa (t) = \Theta (t) \exp (-\gamma t/2) \sin (\tilde\Omega
t)/M\tilde \Omega$, where $ \tilde\Omega =\sqrt
{\Omega^2-\gamma^2/4}$. The case $ \gamma\approx2\Omega $, $
\gamma\ll 1/\tau_0$ is the most interesting for us. Then
\begin{multline}
\langle\langle \phi^3(0)\rangle\rangle\approx \langle \langle
I_{0}^3 \rangle \rangle\lambda^3\int
\frac{d\omega_{1,2,3}}{(2\pi)^3}
\kappa(\omega_1)\kappa(\omega_2)\kappa(\omega_3)\times
\\
\times\delta(\omega_1+\omega_2+\omega_3)=\frac 2 {27}
\frac{\langle \langle I_{0}^3 \rangle \rangle
\lambda^3}{M^3\Omega^4},
\end{multline}
\begin{gather}
\langle\langle \phi^n(0)\rangle\rangle\approx \frac {n!} {n^{n+1}}
\frac{\langle \langle I_{0}^n \rangle
\rangle\lambda^n}{M^n\Omega^{n+1}}, \label{phin}
\end{gather}
where $ \langle \langle I_{0}^n \rangle \rangle $ is the
irreducible current correlator  of the order of $n$, defined by
the quantum binomial distribution [in (\ref{phin}) we neglected
the contribution of the own thermal ammeter noise.]

Thus measuring \textit{irreducible} correlators of coordinate
$\phi$ of the ammeter it is possible to measure irreducible
correlators of currents of high orders in a limit of zero
frequencies (in particular $ \langle \langle I_0^3\rangle
\rangle\propto-T^2 (1-T) $). Such measurements are possible also
under less restrictive  requirements on frequencies of the ammeter if
the kernel $f (x) $ defines the area of an integration so that
$x\ll v_F\tau_0$.  In the opposite limit it is natural to expect
``classical'' answers for the correlators (in particular $ \langle
\langle I_0^3 \rangle \rangle\propto T (1-T) (1-2T) $).

We are grateful to M. Reznikov and especially to D. Ivanov for
fruitful discussions. D.Ivanov paid our attention to special
coordinate and frequency dependence  of the third order
correlators  which appeared in essence important for reviewing
various conditions of measuring. We also are grateful to
M.Feigelman for reading manuscript and useful remarks.

Our work is supported by the Russian Science-Support Foundation,
Russian Foundation  for Basic Research (RFBR), the Russian
ministry of science (the project ``Physics of quantum
computations''), SNF (Switzerland).

\end{document}